\documentclass{aa}  

\usepackage{color}
\usepackage{graphicx}
\usepackage{subcaption}
\usepackage{gensymb}
\usepackage{txfonts}
 \newcommand{\virg}[1]{``#1''}
\begin{document} 
	\title{Multi-wavelength campaign on NCG~7469. 
		IV.\\ The broad-band X-ray spectrum.}
	\author{R. Middei
		\inst{1  \thanks{riccardo.middei@uniroma3.it}}
		\and S. Bianchi 	\inst{1} \and M. Cappi \inst{2}	\and P-O. Petrucci	\inst{3}	\and F. Ursini \inst{2} \and N. Arav	\inst{4,5}	\and E. Behar \inst{4,6} \and G. Branduardi-Raymont\inst{7} \and E. Costantini \inst{8} \and B. De Marco \inst{9} \and L. Di Gesu\inst{10} \and J. Ebrero \inst{11} \and J. Kaastra \inst{8,12} \and S. Kaspi\inst{13} \and G. A. Kriss \inst{14} \and J. Mao\inst{8} \and M. Mehdipour\inst{8} \and  S. Paltani\inst{10} \and U. Peretz \inst{4} \and G. Ponti\inst{15}
	}
	\institute{Dipartimento di Matematica e Fisica, Universit\`a degli Studi Roma Tre, via della Vasca Navale 84, I-00146 Roma, Italy
		\and
		INAF-IASF Bologna, via Gobetti 101, 40129 Bologna, Italy
		\and
		Univ. Grenoble Alpes, CNRS, IPAG, F-38000 Grenoble, France
		\and
		Department of Physics, Technion, Haifa 32000, Israel
		\and
		Department of Physics, Virginia Tech, Blacksburg, VA 24061, USA
		\and
		Department of Astronomy, University of Maryland, College Park, USA
		\and
		Mullard Space Science Laboratory, University College London, Holmbury St. Mary, Dorking, Surrey, RH5 6NT, UK
		\and
		SRON Netherlands Institute for Space Research, Sorbonnelaan 2, 3584 CA Utrecht, The Netherlands
		\and
Nicolaus Copernicus Astronomical Center, Polish Academy of Sciences, Bartycka 18, PL-00-716 Warsaw, Poland
		\and
		Department of Astronomy, University of Geneva, 16 Chemin d'Ecogia, 1290 Versoix, Switzerland
		\and
		European Space Astronomy Centre, PO Box 78, 28691 Villanueva de la Ca\~nada, Madrid, Spain
		\and
		Leiden Observatory, Leiden University, PO Box 9513, 2300 RA Leiden, The Netherlands
		\and
		School of Physics \& Astronomy and Wise Observatory, Raymond and Beverly Sackler Faculty of Exact Sciences, Tel-Aviv
University, Tel-Aviv 69978, Israel
		\and
		Space Telescope Science Institute, 3700 San Martin Drive, Baltimore, MD 21218, USA
		\and
		Max-Planck-Institut f\"ur extraterrestrische Physik, Giessenbachstrasse, 85748 Garching, Germany
		%          \thanks{The university of heaven temporarily does not}
	}

	\abstract{}
	
	\abstract
	% {} leave it empty if necessar@inreference{ID,
	{We conducted a multi-wavelength six-month campaign to observe the Seyfert galaxy NGC~7469, using the space-based observatories \textit{HST}, \textit{Swift}, \textit{XMM-Newton} and \textit{NuSTAR}. Here we report the results of the spectral analysis of the 7 simultaneous \textit{XMM-Newton} and \textit{NuSTAR} observations. The sources shows significant flux variability within each observation, but the average flux is less variable among the different pointings of our campaign. Our spectral analysis reveals a prominent narrow neutral \ion{Fe} K$\alpha$ emission line in all the spectra, with weaker contributions from Fe K$\beta$, neutral Ni K$\alpha$ and ionised iron. We find no evidence for variability or relativistic effects acting on the emission lines, which indicates that they originate from distant material.
    Analysing jointly \textit{XMM-Newton} and \textit{NuSTAR} data a constant photon index is found ($\Gamma$=$1.78\pm0.02$), together with a high energy cut-off $E_{\rm{cut}}=170^{+60}_{-40}$ keV. Adopting a self-consistent Comptonization model, these values correspond to an average coronal electron temperature of kT=$45^{+15}_{-12}$ keV and, assuming a spherical geometry, an optical depth $\tau=2.6\pm0.9$. The reflection component is consistent with being constant, with a reflection fraction in the range $R=0.3-0.6$.
		A prominent soft excess dominates the spectra below 4 keV. This is best fit with a second Comptonization component, arising from a \virg{warm corona} with an average $kT=0.67\pm0.03$ keV and a corresponding optical depth $\tau=9.2\pm0.2$.
	}%
	
	

	%%%%%%%%%%%%%%%%%%%%%%%%%%%%%%%%%%%%%%%%%%%%%%%%%%
	
	%%%%%%%%%%%%%%%%% BODY OF PAPER %%%%%%%%%%%%%%%%%%
	\maketitle
	\section{Introduction}
	
Active galactic nuclei (AGN) are among the brightest sources in the Universe and they account for a large fraction of the X-ray photons we observe in the sky. It is commonly accepted that AGN are powered by matter accreting onto a super massive black hole (SMBH). In the innermost region of the host galaxy, a SMBH is surrounded by a disk of spiralling matter that is responsible for its optical/UV emission, while the physical origin of the higher energetic photons still remains elusive. According to the commonly accepted scenario, X-rays are produced in a region close to the central black hole (BH), the so-called hot corona \citep[e.g.][]{haar91,haar93,haar94}, in which seed optical/UV photons arising from the accreting disk interact with hot thermal electrons through an inverse Compton process. This process reproduces the power law shape we commonly observe in AGN spectra \cite[e.g.][]{Guai99b,Bian09,Mari14}. Moreover, a high energy cut-off is observed in different AGN spectra \citep[e.g][]{Pero02,DeRo02,Guai10,Bren14,Mari14,Lohf15,Ursi16,Alessia17,Porq17,Tort17}, and this is another signature of the thermal Comptonization acting in the hot corona \citep{haar91,haar93}. Furthermore, the primary continuum emission can be modified by a Compton reflection from the disk, from farther material or by absorption from neutral or from ionized gas.
	
As an additional hallmark of AGN activity there is continuum variability. Flux variations are indeed observed on several time-scales, from hours and days \cite[e.g.][]{Pont12} up to years and decades \citep[e.g.][]{Vagn11,Vagn16,Midd17}. Rapid variability not only is of primary importance to investigate the X-ray emission, but it can be also used to estimate the SMBH mass \citep[e.g.][]{McHa06,Pont12} and as a luminosity distance estimator \citep{LaFa14}.

Long multi-wavelength monitorings of single nearby AGN produced outstanding results \citep[e.g.][and the related series of papers on Mrk~509]{Kaas11}. The target of our observational campaign is NGC~7469, a luminous Seyfert galaxy \citep[$L_{\rm{bol}}\sim10^{45}$ erg~s$^{-1}$,][]{Bear17} at $z=$0.016268 \citep{Spri05}. Using reverberation mapping, \citet{Pete14} found that NGC~7469 hosts a BH with a mass of $1.1\pm0.1\times10^7$ $M_{\sun}$ and an Eddington ratio of the order of 0.3. In the X-rays, this Seyfert galaxy was first observed by the \textit{Uhuru} satellite \citep{Forman78} in the seventies, and it was subsequently studied by many other observatories that found this source to have a complex X-ray emission. Since the \textit{EXOSAT} observation, we know that its X-ray spectrum displays an excess in the soft band \citep{Barr86}. Other authors \citep{Turn91,Bran93,Guai94,Nand98,Nand00,DeRo02} analysed this source using data obtained by \textit{Einstein}, \textit{ROSAT}, \textit{ASCA}, \textit{RXTE} and \textit{BeppoSax}.
%	{{\color{red}More recent space-based telescopes such as  \textit{SUZAKU} and \textit{CHANDRA} observed this source. In the work by \cite}}
NGC~7469 was studied also more recently: \cite{Petr04} investigated the UV/X-ray variability, \cite{Scot05} analysed its simultaneous X-ray, far-ultraviolet, and near-ultraviolet spectra using \textit{Chandra}, \textit{FUSE} and \textit{STIS}, while \cite{Patr11} studied this source taking advantage of \textit{Suzaku} observations. Previous \textit{XMM-Newton} data were analysed in \cite{Blus03} and \cite{DeMa09}, while some results from the 2015 observational campaign have been presented in \cite{Bear17, Peretz17}.

This paper focuses on the 7 simultaneous \textit{XMM-Newton} and \textit{NuSTAR} observations of our campaign, and it is organised as follows: Sect. 2 focuses the NGC 7469 temporal analysis, Sect. 3 focuses on the data reduction. Sect. 4 and Sect. 5 report on the spectral analysis. Sect. 6 contains the discussion of the results, and in Sect. 7 a summary of this work is reported.

	%The geometrical and physical properties of the X-ray-emitting region are not yet fully understood. On the one hand, the power law-like continuum emission in hard X-rays is generally believed to originate in a hot and compact corona, likely located in the inner part of the accretion flow \citep[e.g.][]{Reis13}. However, an excess of emission below 1-2 keV above the extrapolated high-energy power law is commonly observed in the spectra of AGN \citep[e.g][]{Walt93, Bian09}. Albeit the nature of this so-called soft X-ray excess is uncertain \citep[e.g.][]{Done12}, a possible explanation is thermal Comptonization by a warm, optically thick medium \citep[e.g.][]{Magd98, Petr13}.
	%In particular, \cite{Petr13} studied the high-energy spectrum of  Mrk 509 in detail, using the data from a long, multiwavelength campaign \citep{Kaas11}. This source showed  a correlation between the optical/UV and soft (< 0.5 keV) X-ray flux, but no correlation between the optical/UV and hard (> 3 keV) X-ray flux \citep{Mehd11}.  Indeed, \cite{Petr13} found that the spectrum is well described by a two-corona model: a warm ($kT$$\sim$1 keV), optically thick ($\tau\sim15$) corona responsible for both the optical/UV emission and the soft X-ray excess, and  a hot ($kT$$\sim100$ keV), optically thin ($\tau\sim0.5$) corona responsible for the hard X-ray emission.

	% \citep[e.g.][]{Author2012}.
	
	\section{Observations and data reduction}
	
The spectral analysis presented in this work is based on \textit{XMM-Newton} \citep{Jans01} and \textit{NuSTAR} \citep{Harr13} observations of NGC~7469 belonging to the multi-wavelength campaign first described by \citet{Bear17}. The two satellites observed the source simultaneously between June 12 and December 28, 2015. The 7 observations are spaced by different time intervals, allowing us to study flux and spectral variations on different time-scales, see Tab. 1.
	
	\begin{table}
			\caption{For the seven observations analysed in this work we report for each satellite the observation ID, the start date and the net exposure time (ks).}
		\begin{tabular}{c c c c}
			\hline
			Obs. Satellites & Obs. ID & Start Date & Net Exp. (ks) \\
			\hline
			\hline
			\textit{XMM-Newton} & 0760350201 & 2015-06-12 & 63\\
			\textit{NuSTAR} & 60101001002 & 2015-06-12 & 21 \\
			\hline
			\textit{XMM-Newton} & 0760350301 & 2015-11-24 & 59\\
			\textit{NuSTAR} & 60101001004 & 2015-11-24  & 20 \\
			\hline
			\textit{XMM-Newton} & 0760350401 & 2015-12-15& 59 \\
			\textit{NuSTAR} & 60101001006 & 2015-12-15 & 22 \\
			\hline
			\textit{XMM-Newton} &0760350501  & 2015-12-23 & 62 \\
			\textit{NuSTAR} & 60101001008 & 2015-12-22 & 23 \\
			\hline
			\textit{XMM-Newton} & 0760350601 & 2015-12-24 & 65 \\
			\textit{NuSTAR} & 60101001010& 2015-12-25 & 21 \\
			\hline
			\textit{XMM-Newton} & 0760350701 & 2015-12-26 & 67 \\
			\textit{NuSTAR} &60101001012 & 2015-12-27 &21  \\
			\hline
			\textit{XMM-Newton} & 0760350801 & 2015-12-28 & 70 \\
			\textit{NuSTAR} &60101001014  & 2015-12-28 & 23\\
			\hline
		\end{tabular}
	\end{table}

\textit{XMM-Newton} data were obtained using the EPIC cameras \citep{Stru01,Turn01} in the Small Window operating mode and they were processed taking advantage of the \textit{XMM-Newton} Science Analysis System\footnote{"Users Guide to the XMM-Newton Science Analysis System", Issue 13.0, 2017 (ESA: XMM-Newton SOC).} ($SAS$, Version 15.0.0). Because of its larger effective area with respect to the two MOS cameras, we only report the results for the PN instrument. We extract spectra from circular regions of 50 arcsec radius for the background, and 40 arcsec radius for the source.  These regions are selected by an iterative process that maximizes the signal-to-noise ratio \citep{Pico04}. All the spectra were rebinned in order to have at least 30 counts for each bin and not to oversample the spectral resolution by a factor greater than 3.

\textit{NuSTAR} data were reduced taking advantage of the standard pipeline ($nupipeline$) in the \textit{NuSTAR} Data Analysis Software (nustardas release: nustardas\_14Apr16\_v1.6.0, part of the \textit{heasoft} distribution\footnote{NuSTARDAS
	software guide, Perri et al. (2013), https://heasarc.gsfc.nasa.gov/docs/nustar/analysis/nustar\_swguide.pdf}), adopting the latest calibration database. The \textit{NuSTAR} observatory carries in its focal plane two modules A and B corresponding to the hard X-ray detectors FPMA and FPMB. Spectra and light curves were extracted for both modules using the standard tool \textit{nuproducts}. A circular region with radius of $\sim$70 arcsec is used to extract the source counts while the  background is obtained from a blank area with the same radius, close to the source. Similarly to the \textit{XMM-Newton} spectra, we have binned \textit{NuSTAR} spectra in order to  have  a  signal-to-noise ratio  greater  than  5  in  each  spectral  channel,  and to not oversample the instrumental resolution by a factor greater than 2.5. The spectra of the two modules are in good agreement with each other, the cross-normalization constant for FPMB with respect to FPMA in all the fits being unity within 1 per cent. 
    Spectra were analysed with \textit{XSPEC} 12.9 \cite{Arna96}. 
    
    \indent All errors reported in the plots account for 1$\sigma$ uncertainty, while errors in text and tables are quoted at 90\% confidence level, unless otherwise stated.
	\section{Temporal analysis}
\begin{figure*}
		\centering
		\includegraphics [width=7.0in]{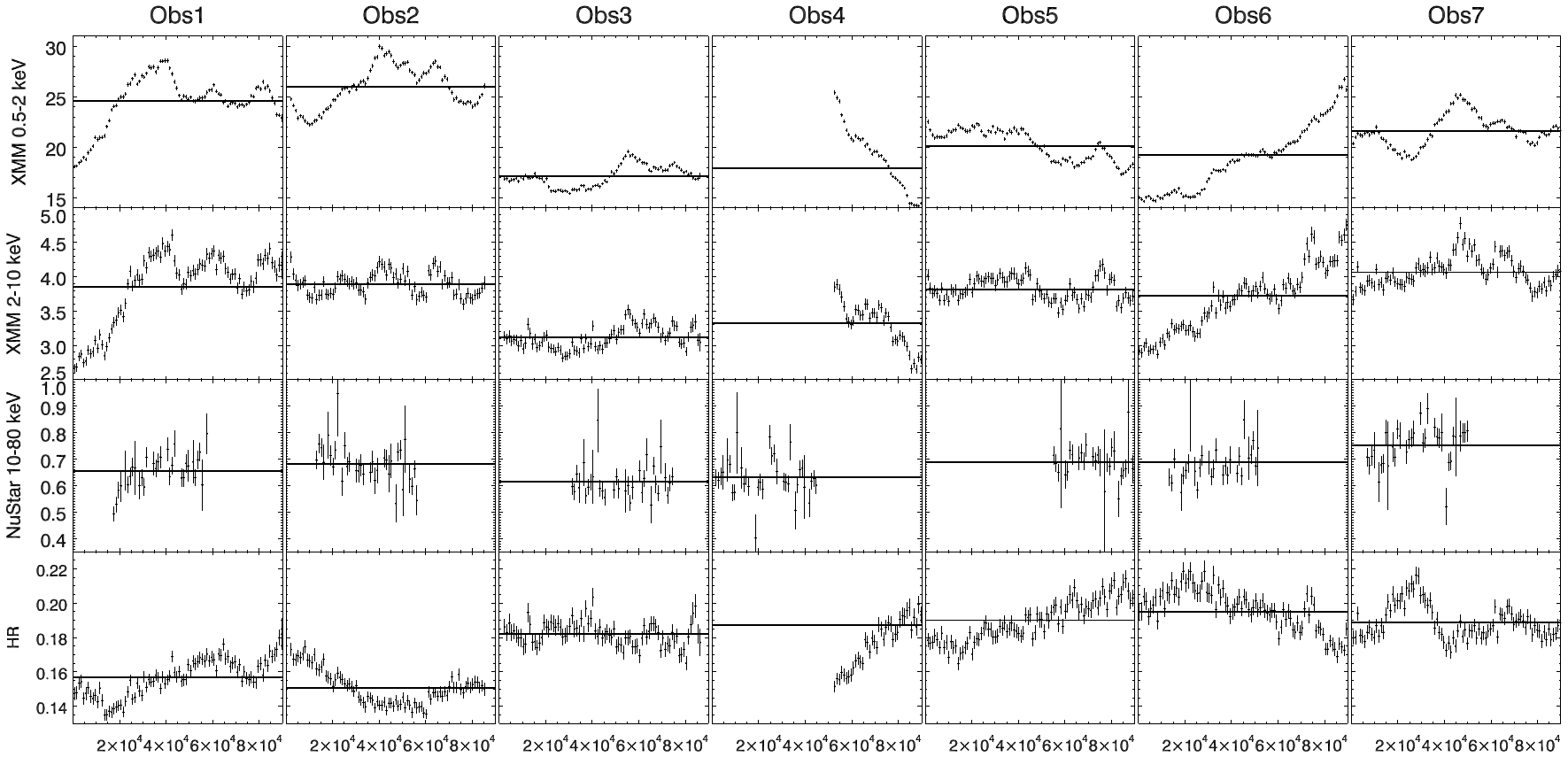}
		\caption{\label{lightcurves}Light-curves corresponding to the 7 simultaneous observations performed by \textit{XMM-Newton} and \textit{NuSTAR} for NGC~7469, in three different energy bands: 0.5-2 keV (first row), 2.0-10 keV (second row), 10-80 keV (third row). Panels in the fourth row show the ratio between the soft 0.5-2 keV and the hard 2.0-10 keV light-curves. The axis units are sec and counts respectively for the x-axis and the y-axis. The solid line overlapping all the light-curves in each sub-panel represents the average of the counts for the specific light-curve.}
	\end{figure*}

	We started investigating the NGC 7469 temporal properties computing the light-curves for all the observations. We used the standard command \textit{epiclccorr} for the \textit{XMM-Newton} data to compute corrected for the background light-curves in the 0.5-2 keV and 2-10 keV bands, while, we used the \textit{nuproducts} pipeline to compute the \textit{NuSTAR} light-curves in the 10-80 keV band.
	The \textit{XMM-Newton} and \textit{NuSTAR} light-curves of all the observations of our campaign are shown in Fig~\ref{lightcurves}.  The source shows a remarkable intra-observation flux variability (e.g. up to $\sim$60\% in the sixth observation for the \textit{XMM-Newton} in the 0.5-2 and 2-10 keV bands), with significant flux variations on time scales of a few ks. On longer time-scales, flux variability appears to be less significant. The mean counts for each of the 7 observations is observed to be, in average, weakly variable ($\sim$15\%, $\sim$9\% and few per cent in the 0.5-2, 2-10 and 10-80 keV bands, respectively). 
	
	A convenient analysis tool for variability characterization is the so-called “normalized excess variance” (NXS). The NXS is defined as $\sigma^2_{rms}=\frac{1}{N\mu^2}\sum_{i=1}^{N}(X_i-\mu)^2-\sigma_i^2$~, where $\mu$ is the unweighted count rate mean within the segment of the light curve, N is the number of the good time bins in that segment, and $X_i$ represents the count rate with $\sigma_i^2$ as associated uncertainty. We computed the $\sigma^2_{rms}$ in the 2-10 keV band for all the observations of our campaign in the 20 ks time bin (see \cite{Pont12} for more details), finding an average value $\sigma^2_{rms}$=0.0021$\pm$0.0005. A tight correlation between  the X-ray variability of the source and its BH mass has been found by several authors \citep[e.g.][]{Nand97,Vaug03,McHa06,Pont12}. In particular, adopting the relation among $\sigma^2_{rms}$ and $M_{BH}$ in \cite{Pont12}, we are able to estimate the BH mass $M_{BH}= 1.1\pm0.1\times10^{7}$ $M_{\sun}$ for NGC~7469, in very good agreement with the reverberation mapping estimate by \cite{Pete14}.
	The soft X-ray appears to be the most variable band, but hardness ratios do not display a large variation (however significant from a statistical point of view, see Sect. 6) both within and among the observations ($\sim$8\%), see Fig. 1.\\
	Therefore, we decided to use the average spectra of each observation in the following spectral analysis.
	\begin{figure*}[h!]
	\includegraphics[width=0.5\textwidth, height=0.33\textwidth]{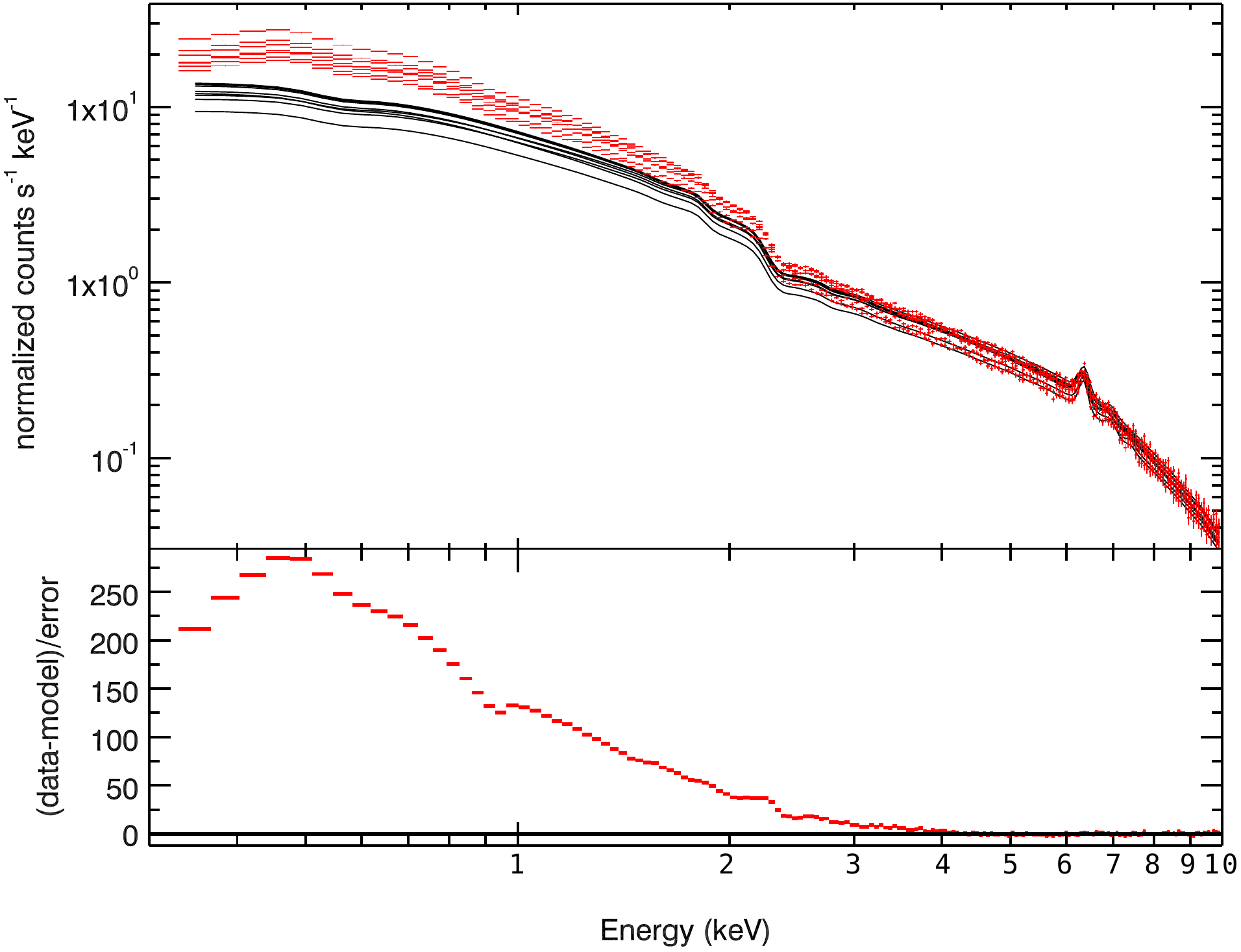}	
	\includegraphics[width=0.5\textwidth]{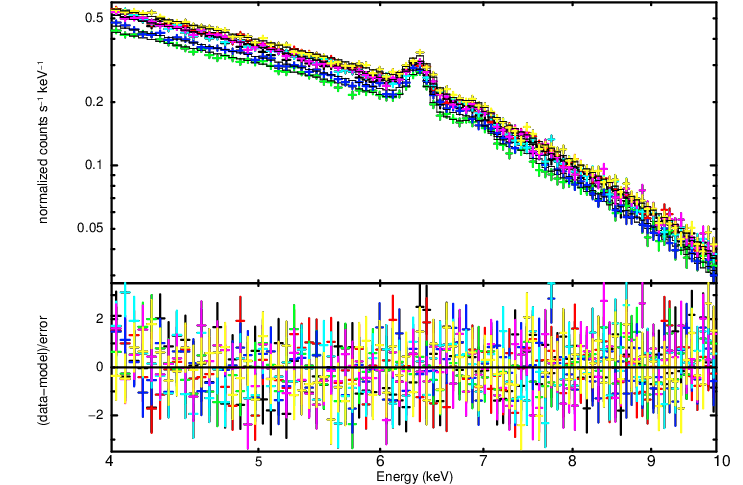}
	
	\includegraphics[width=0.5\textwidth]{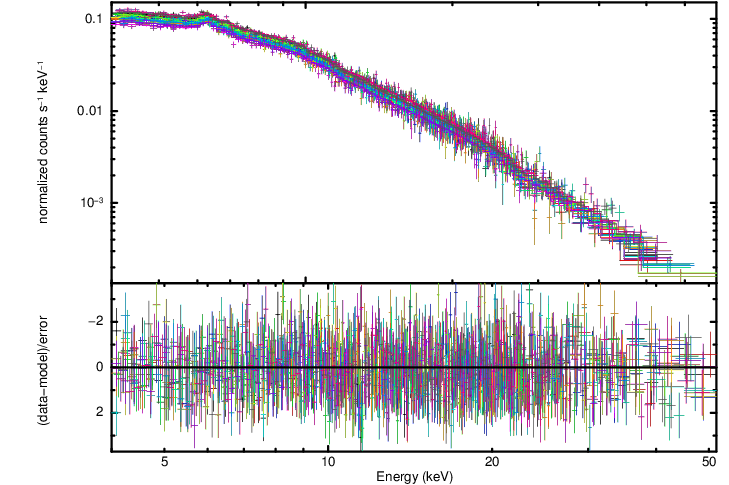}	
	\includegraphics[width=0.5\textwidth]{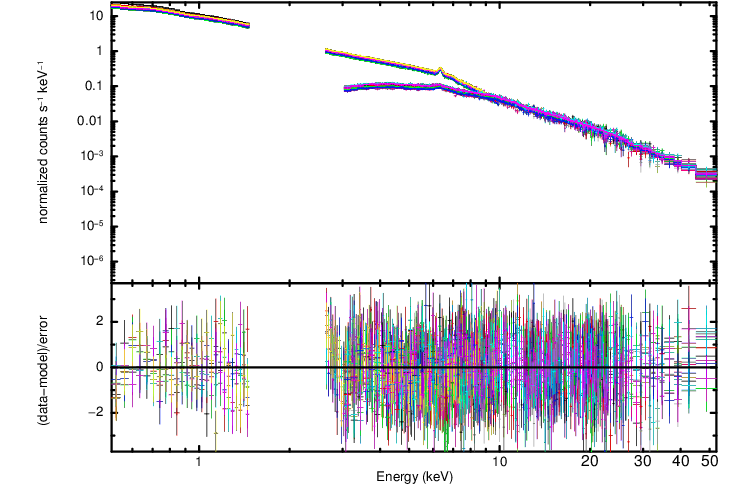}	
	\caption{$\;$ In the top left hand panel the 7 \textit{XMM-Newton} spectra are displayed in red, while, in black, the best fit model to the data above 4 keV. In the bottom part of the plot we report the residuals of the 7 spectra that are grouped for plotting purposes using in \textit{Xspec} the command "setplot group". A secondary component clearly extends up to 4 keV. $\;$ In the top right side plot the best fit to the 7 \textit{XMM-Newton} data and the corresponding residuals with the basic model (i.e. \textit{zashift$\times$(pexrav+zgauss+zgauss))}, at which we added other Gaussian lines when weaker emission lines were found in the observation.	$\;$ The bottom left side plot shows the best fit for all \textit{NuSTAR} spectra above 4 keV. In its top sub-panel the best fit obtained adopting the model that employs \textit{xillver} (see Sect. 4.2), while the sub-bottom one refers to the corresponding grouped residuals. $\;$ In the last panel, the best fit ($\chi^2$=3041 for 2765 d.o.f.) for the NGC~7469 \textit{XMM-Newton} and \textit{NuSTAR} spectra, see Sect. 5.1 for the model details.}
	
\end{figure*}
	
	\section{Spectral Analysis: data above 4 keV}

	\subsection{\textit{XMM-Newton}: the iron line energy band}

\begin{figure}[h!]
	\centering
	\includegraphics [width=3.5in]{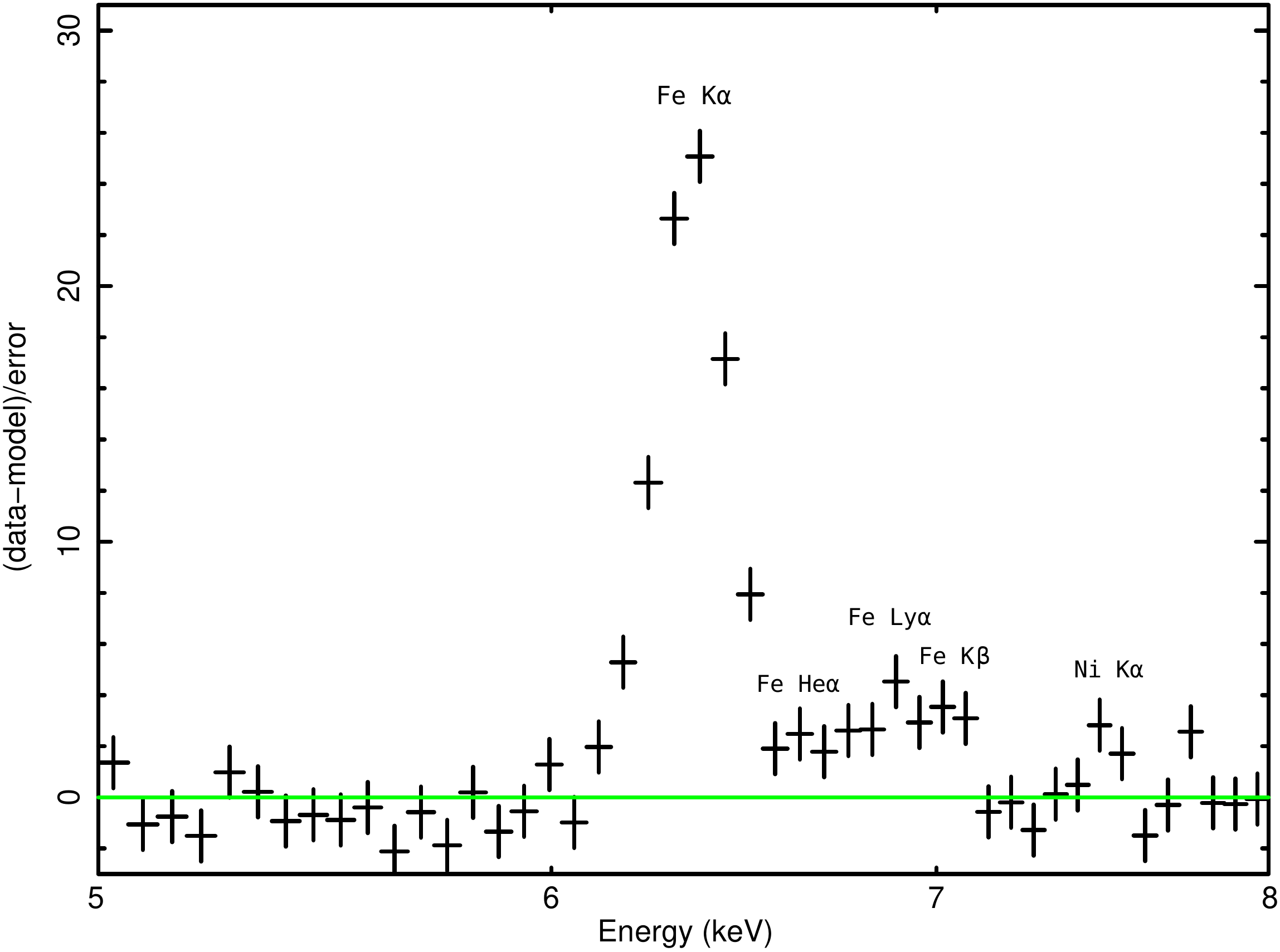}
	\caption{\label{iron}The residuals (in the observer frame) to a simple power law fit to the \textit{XMM-Newton} spectra in the energy range 5-8 keV. Emission lines are clearly present. The residuals of the 7 spectra are grouped for plotting purposes.}
\end{figure}

We start our spectral analysis adopting a simple power law model for all our \textit{XMM-Newton} data. This crude fit leaves strong residuals in the soft X-rays. This \virg{soft-excess} extends up to 4 keV and is showed in Fig. 2, top left hand panel, that is obtained  fitting the data above 4 keV and then plotting the best fit model with the whole spectrum, i.e. extending the model into the soft X-ray band. Therefore, as a first step for our analysis, we decided to characterize the limited energy band 4-10 keV. The residuals in this energy band are dominated by strong emission features, readily identified with the K$\alpha$ lines from neutral and H-like iron, as well as possible presence of weaker contributions from neutral Nickel, Fe K$\beta$ and ionized iron (see Fig.~\ref{iron}).

We therefore adopted the following model to fit the 4-10 keV spectra: \textit{zashift$\times$(pexrav+zgauss+zgauss)}. The  \textit{pexrav} code \citep{Magd95} is adopted to model the primary emission as well as any reflected component likely associated with the fluorescent emission line from neutral iron, model by the first Gaussian line, while the second Gaussian line accounts for the \ion{Fe}{xxvi} Ly$\alpha$  at 6.966 keV. The \textit{zashift} component is used to correct the well known \textit{XMM-Newton} calibration issue affecting the EPIC pn \citep[a detailed discussion on this topic can be found in][]{Capp16}. Within this paper, a \textit{zashift} correction of about $-8\times10^{-3}$ (corresponding to $\sim50$ eV at 6.4 keV), will be always applied for all the \textit{XMM-Newton} spectra. 
To model the data, we let free to vary among the observations the normalizations of both the \ion{Fe} K$\alpha$ and the \ion{Fe}{xxvi} Ly$\alpha$ lines, the reflection parameter and the photon index in \textit{pexrav} as well as its normalization. The iron abundance is also free to vary but tied among the different observations.
This simple model leads us to a very good best fit in the 4-10 keV band ($\chi^2=586$ for 567 $d.o.f.$~), see Fig. 2, top right hand side panel.
	%retrieving the values for the parameters reported in Tab. 2.
	
\begin{figure}
		\centering
		\includegraphics [width=3.0in]{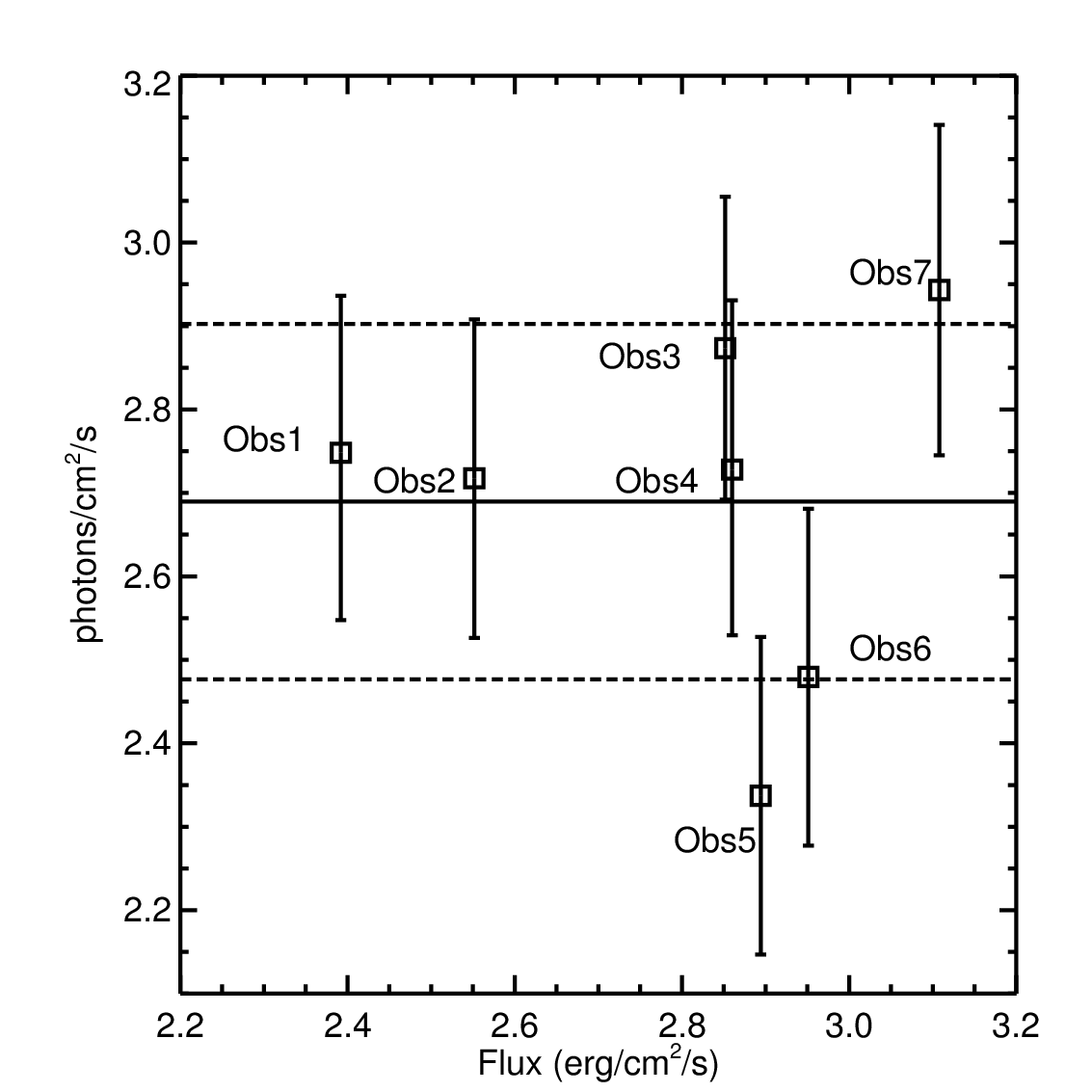}
		\caption{\label{iron_lc} The \ion{Fe} K$\alpha$ line flux vs. 2-10 continuum flux for the 7 \textit{XMM-Newton} observations. The average line flux is shown together with the standard deviation (dashed lines).}
	\end{figure}

The \ion{Fe} K$\alpha$ and the \ion{Fe}{xxvi} Ly$\alpha$ lines are both statistically significant ($>99.9$ per cent according to the F-test), and have a constant flux (3$\sigma$ level) during the 7 observations of our campaign (however \citet{DeMa09} analysing previous \textit{XMM-Newton} data found possible hints of variability in the \ion{Fe}{xxvi} Ly$\alpha$ component), with average values of $3\times10^{-5}$ and $4\times10^{-6}$ ph cm$^{-2}$ s$^{-1}$, respectively, corresponding to equivalent widths of $\sim90$ and $\sim20$ eV (see Fig.\ref{iron_lc} for the \ion{Fe} K$\alpha$ line). Moreover, both lines are narrow, their intrinsic line width being consistent with zero in all the observations. In particular, the line profiles do not show any evidence for the relativistic effects (see Fig.~\ref{iron}) expected to occur in the innermost regions of the accretion disc.

The residuals in Fig.~\ref{iron} suggest the presence of the \ion{Fe}{xxv} He$\alpha$, Fe K$\beta$, and the Ni K$\alpha$ emission lines expected, respectively at 6.64-6.7, 7.06 and 7.47 keV, so we tried to fit them adding three further Gaussian lines to our best fit model. However, the inclusion of these lines does not improve significantly the $\chi^2$ of the fit. In fact, these three lines are very weak: the Fe K$\beta$ flux is only an upper limit in all the spectra, the Nickel line is a detection only in two observations, with an average flux of $\sim 4.6\times10^{-6}$ ph cm$^{-2}$ s$^{-1}$, and \ion{Fe}{xxv} He$\alpha$ is detected just in three observations with an average flux
$\sim 4.0\times10^{-6}$ ph cm$^{-2}$. Indeed, none of these lines is more significant than 98 per cent, according to the F-test. The best fit normalizations, or their upper limits, are reported in Table 2 for all the emission lines discussed in this section.

\begin{table}
	\setlength{\tabcolsep}{0.7pt}
		\caption{The best fit normalizations for the emission lines observed in the \textit{XMM-Newton} observations. The normalizations are in units of  $10^{-6}$ ph cm$^{-2}$.}
				\centering
		\begin{tabular}{ c c c c c c c} 
			Obs &\ion{Fe}~K$\alpha$&\ion{Fe}{xxvi}~Ly$\alpha$&~~\ion{Fe}{xxv}~He$\alpha$&~~~Fe~K$\beta$&~~~ Ni~K$\alpha$ \\
			\hline
			\hline
			\\
			1 &$28.7^{+1.8}_{-1.8}$&$4.3^{+1.6}_{-1.7}$&<3.12&<5.2&<3.3\\
			\\
			2 &$24.7^{+2.0}_{-2.0}$&$3.5^{+2.0}_{-2.0}$&<1.7&<6.3& $4.3^{+2.8}_{-3.0}$\\
			\\
			3 &$27.4^{+2.0}_{-1.9}$&$6.1^{+2.0}_{-1.7}$&<3.1&<6.9&<4.21\\
			\\
			4 &$27.1^{+1.9}_{-1.9}$&$2.6^{+1.9}_{-1.9}$&$4.2^{+2.8}_{-2.5}$&<4.4&$4.9^{+3.0}_{-3.0}$\\
			\\
			5 &$23.3^{+1.9}_{-1.9}$&$2.4^{+2.0}_{-2.0}$&$4.1^{+3.0}_{-2.7}$&<6.0& <4.9\\
			\\
			6 &$27.2^{+2.0}_{-1.9}$&$4.4^{+2.0}_{-2.0}$&<4.2&<6.9&<4.0\\
			\\
			7 &$29.4^{+2.0}_{-1.9}$&$4.7^{+2.0}_{-2.0}$&$3.9^{+2.8}_{-2.8}$&<5.5& <6.4\\
			\hline
			
		\end{tabular}
	\end{table}
	\subsection{\textit{NuSTAR}: the 4-80 keV spectra}
	
Our preliminary analysis on the 4-10 keV \textit{XMM-Newton} data shows a non variable and narrow neutral iron line, likely produced by Compton-thick gas far from the central SMBH. For the \textit{NuSTAR} data we adopted the self-consistent model \textit{xillver} \citep[][]{Garc10,Garc13}, to fit both the neutral iron line and the associated reflection continuum (we use \textit{xillver} with its ionization parameter fixed to zero). A Gaussian emission line is included to model the observed emission line at 6.966 keV, and a second Gaussian line is used for the \ion{Fe}{xxv} He$\alpha$ lines. Strong soft-excess features appear also in \textit{NuSTAR} spectra, thus, similarly to what we already performed for \textit{XMM-Newton}, we preliminary cut \textit{NuSTAR} spectra below 4 keV. The reflection fraction, the high energy cut-off, the normalizations and the inter-calibration constant are free to vary, while the iron abundance is free to vary but tied among the observations. Following this approach, we obtained the best fit to the data ($\chi^2=1762$ for 1665 d.o.f.), plotted in Fig. 2, bottom left hand panel. The values of the best fit parameters are reported in Tab. 3.

	\begin{table}
		\setlength{\tabcolsep}{0.4pt}
		\centering 
				\caption{The values from the best-fit model for all the \textit{NuSTAR} spectra analysed in the 4-78 keV band using the best-fit model \textit{xillver+zgauss+zgauss} ($\chi^2=1762$ for 1665 d.o.f.). The 10-78 keV flux is in units of $10^{-11}$ erg/cm$^2$/s, and the iron abundance was let free to vary but tied among the observations.}
		\begin{tabular}{c c c c c c c}
			%nustarallobs_sb1zgausserr90.txt e poi
			\hline	 
			\hline
			Obs &  $\Gamma$ & $E_{cut} (keV)$ & R & Norm$_{\rm{xi}}$($10^{-4}$)& A$^\dag$$_{\rm{Fe}}$&Flux$_{10-78}$\\
			\hline
			\hline
			\\
			1  & $1.82^{+0.06}_{-0.06}$ & $110^{+85}_{-35}$ & $0.50^{+0.14}_{-0.10}$&$1.56^{+0.10}_{-0.08}$&2.8$\pm$0.6&$7.4^{+0.4}_{-0.3}$ \\
			\\
			2  & $1.77^{+0.06}_{-0.06}$ & $190^{+650}_{-90}$ & $0.32^{+0.11}_{-0.10}$&$1.84^{+0.07}_{-0.30}$&2.8$\pm$0.6&$8.2^{+0.7}_{-0.4}$\\
			\\
			3  & $1.73^{+0.07}_{-0.07}$ & $85^{+60}_{-20}$ & $0.62^{+0.16}_{-0.13}$&$1.23^{+0.09}_{-0.07}$&2.8$\pm$0.6&$6.6^{+0.4}_{-0.3}$ \\
			\\
			4  & $1.83^{+0.03}_{-0.05}$ & $>230$ & $0.33^{+0.08}_{-0.09}$&$2.14^{+0.02}_{-0.29}$&2.8$\pm$0.6&$7.6^{+0.6}_{-0.6}$\\
			\\
			5  & $1.78^{+0.05}_{-0.06}$ & $>120$ & $0.34^{+0.10}_{-0.10}$&$1.89^{+0.30}_{-0.20}$&2.8$\pm$0.6&$8.1^{+0.6}_{-0.3}$ \\
			\\
			6  & $1.75^{+0.06}_{-0.06}$ & $>110$ & $0.37^{+0.11}_{-0.10}$&$1.75^{+0.30}_{-0.19}$&2.8$\pm$0.6&$7.8^{+0.8}_{-0.6}$\\
			\\
			7  & $1.78^{+0.05}_{-0.05}$ & $195^{+420}_{-80}$ & $0.35^{+0.10}_{-0.09}$&$1.98^{+0.30}_{-0.16}$&2.8$\pm$0.6&$8.9^{+0.6}_{-0.4}$\\
			\hline
		\end{tabular}

	\end{table}
	%     7      2.18298      3.50382    (-0.60087,0.719964) from nustarallobs_sb1.xcm
	
	This best fit model requires a super-Solar iron abundance $A_{\rm{Fe}}=2.8\pm0.6$. The photon index is consistent with being constant between the observations, with an average value of 1.78$\pm$0.02 (see Fig. 5, top left side panel, showing contour plots ). A high energy cut-off is well constrained only in two observations, with values around 150 keV, with some indications of variability up to larger values in other observations (see same panel in Fig. 5). However, fitting the \textit{NuSTAR} spectra tying the high energy cut-off among the 7 observations yields a value of $170_{-40}^{+60}$ keV with a $\chi^2$=1775 for 1671 d.o.f. very similar to the previous one.
	Some hints of variability are also found for the reflection fraction in the range 0.3-0.6 (see Fig. 5, top right hand side plot, showing contour plots), with the flux of the reflection component consistent with being constant when the primary
	continuum varies (we obtained larger reflection components for lower flux states), in agreement with an origin from distant matter.
	
	Although the iron K$\alpha$ line does not present any broadening of its profile, we tested for the presence of a relativistic reflection component. Therefore, we added to the best-fit model a further reflection component, accounting for the relativistic effects arising in matter in the innermost regions of the accretion disk. We used \textit{relxill} \citep[e.g.][]{Garc14a,Daus16}. The photon index and the high energy cut-off are tied between \textit{relxill} and \textit{xillver}, while the \textit{relxill} ionization parameter $\xi$ and its normalization are free to vary in all the observations. No significant improvement in terms of $\chi^2/d.o.f.$ is found: $\Delta\chi^2$= 30 for 18 d.o.f. less, corresponding to 80 per cent confidence level according to F-test. No relativistic reflection is required by the \textit{NuSTAR} data: indeed, in all but three of the observations, the normalization of \textit{relxill} is consistent with zero.

	According to the standard scenario, hard X-rays are produced by Comptonization, thus as a further investigation, we decided to model the \textit{NuSTAR} spectra using a self-consistent Comptonization model. In our model we substitute  \textit{xillver} with \textit{xillvercp} \citep[][]{Garc14a,Daus16}. This different code accounts for the primary emission produced by \textit{nthcomp}, \citep{Zdzi96}, and a non-relativistic reflection. The parameters of this new model are treated as in the previous fit, and the hot corona electron temperature is free to vary.
	The best fit obtained ($\chi^2=1768$ for 1658 d.o.f.) adopting \textit{xillvercp} shows larger values for the photon index with respect to the previous best-fit model with \textit{xillver} (on average 0.08). On the other hand, the parameters of the reflection component are in agreement with the values previously quoted.
    The best-fit values for the parameters of this fit are displayed in Table 4~.
    	\begin{table}
		\setlength{\tabcolsep}{0.4pt}
		\centering 
				\caption{Values and errors for the best-fit parameters of all the \textit{NuSTAR} spectra analysed in the 4-78 keV band adopting \textit{xillvercp}. Optical depth estimates are obtained following \cite{Belo99}, see Sect. 5.1 for details. The iron abundance is free to vary but tied among the observations.}
		\begin{tabular}{c c c c c c c}
			\hline	 
			\hline
			Obs &  $\Gamma_{\rm{hard}}$ & $kT_{\rm{hc}} (keV)$ &R& $\tau_{\rm{hc}}$&Norm$_{\rm{xicp}}$ ($10^{-4}$)&A$^\dag$$_{\rm{Fe}}$\\
			\hline
			\hline
			\\
            	1  & $1.89^{+0.02}_{-0.02}$ & $20^{+15}_{-4}$ &$0.45^{+0.10}_{-0.10}$&$3.0^{+0.1}_{-0.1}$&$1.39^{+0.18}_{-0.11}$&2.4$\pm$0.4\\
			\\
			2  & $1.83^{+0.02}_{-0.01}$ & $>21$ &$0.27^{+0.10}_{-0.09}$& $>3.1 $&$1.75^{+0.11}_{-0.30}$&2.4$\pm$0.4\\
			\\
			3  & $1.85^{+0.02}_{-0.02}$ & $22^{+16}_{-5}$ &$0.52^{+0.11}_{-0.11}$& $2.9^{+0.1}_{-0.1}$ &$1.25^{+0.11}_{-0.09}$&2.4$\pm$0.4\\
			\\
			4  & $1.85^{+0.01}_{-0.02}$ & $>20$ &$0.30^{+0.10}_{-0.11}$& $>3.3$ &$2.17^{+0.05}_{-0.30}$&2.4$\pm$0.4\\
			\\
			5  & $1.82^{+0.01}_{-0.02}$ & $>20$ &$0.30^{+0.09}_{-0.09}$& $>3.2$&$1.95^{+0.20}_{-0.19}$&2.4$\pm$0.4 \\
			\\
			6  & $1.81^{+0.01}_{-0.02}$ & $>26$ &$0.29^{+0.10}_{-0.09}$& $>2.8$ &$1.98^{+0.30}_{-0.18}$&2.4$\pm$0.4\\
			\\
			7  & $1.83^{+0.01}_{-0.01}$ & $37^{+140}_{-15}$ &$0.30^{+0.10}_{-0.10}$& $2.2^{+4.0}_{-0.1}$&$1.94^{+0.20}_{-0.17}$&2.4$\pm$0.4 \\

			\hline
		\end{tabular}
	\end{table}

	In Fig. 5, bottom left hand panel, the contour plots for the photon index and the hot electron temperature are reported. Weak variations in the photon index are observed, while the electron temperature has a more constant behaviour.  
	To test the variability of the hot electron temperature, similarly to what we already performed for the high energy cut-off, we fit the \textit{NuSTAR} spectra tying the electron temperature among the observations of this campaign. This yielded a measure of kT=$45^{+15}_{-12}$ keV corresponding to a best fit ($\chi^2=1773$ for 1664 d.o.f.~) very similar to the previous one.

	\section{Spectral Analysis: 0.5-80 keV band}
	\subsection{\textit{XMM-Newton}+\textit{NuSTAR}: broad-band spectrum }	
	As shown in Sect. 4.1 (see also Fig. 2, top left hand panel) the data below 4 keV are characterized by a strong soft-excess. 
	In order to properly model the continuum emission associated with this excess, we first characterize any discrete emitting and absorbing feature expected in the 0.5-4 keV band. On one side, these are features that can be directly attributed to the detector systematic calibration uncertainties, i.e. issues on its quantum efficiency at the Si K-edge (1.84 keV), and on the mirrors effective area at the Au M-edge ($\sim 2.4$ keV). To avoid these issues, we ignored the spectral bins in the energy range 1.7-2.6 keV \citep[see e.g. ][]{Kaas11,DiGe15,Ursi15,Capp16}. We then included all the emission and absorption features (e.g due to the warm absorbers) derived from the analysis of the \textit{XMM-Newton} \textit{RGS} spectra of our campaign by \citet{Bear17}. Since none of these components display significant variability during our campaign \citep{Bear17,Peretz17}, we keep all their parameters fixed in our following fits to the values found from the \textit{RGS} data \citep[see][for a detailed description of all the components]{Bear17}. Some line-like features still remained in the \textit{pn} spectra, and even if very weak, they result significant in terms of $\chi^2$, as a consequence of the high number of counts in the soft band. Two narrow Gaussian lines untied and free to vary among the observations at $\sim$0.75 keV and $\sim$1 keV are enough to correct these residual narrow features \citep[see also e.g.][]{Kaas11,DiGe15,Ursi15,Capp16}.
	
At first, we model the soft X-ray emission with a phenomenological continuum model, such as a power law or a black body. However, in both cases we do not get an acceptable fit, obtaining $\chi^2$=3542 for 2793 d.o.f. and $\chi^2$=7430 for 2793 d.o.f. respectively. We then tried to reproduce the soft excess via two self-consistent models: blurred relativistic reflection arising from the innermost regions of the accretion disk, and Comptonization from a warm corona.

There are a number of reasons (e.g. high BH spin, high ionization parameters) that could lead to a weak broad iron line, but still a prominent relativistic reflection continuum, particularly in the soft X-rays. Indeed, \citet{Walt13}, analysing \textit{Suzaku} data, found a good fit modelling the NGC 7469 soft excess using a relativistic reflection model. Thus, even if our previous analysis failed in finding significant signatures from relativistic effects in the hard X-ray band (and, notably, in the iron line profile), we tested for a relativistic origin for the soft excess in this source.	

To perform this test, we again added \textit{relxill} to the model, similarly to what previously done in the \textit{NuSTAR} data alone, leaving the ionization parameter, the coronal emissivity, the black hole spin and the normalization free to vary among the observations, while the photon index and the high energy cut-off are linked to those in \textit{xillver}. We get parameters ($\xi$=2.4$\pm$0.2, i=45$^\circ$, emissivity=4.8$\pm$0.3, a>0.996) consistent with those found by \citet{Walt13}, but our fit is not statistically acceptable ($\chi^2$=6036 for 2788 d.o.f.). This discrepancy is likely due to the much higher S/N of our data (especially in the soft X-rays) with respect to that used by \citet{Walt13}.

We finally tried \textit{nthcomp} \citep{Zdzi96,Zyck99}, accounting for a Comptonized continuum from a warm corona, as discussed by \citet{Petr13,Roza15,Petr17}. For this model we untie and let free to vary the electron temperature and the seed photons temperatures among the observations.
For each observation, all the parameters among \textit{XMM-Newton} and \textit{NuSTAR} are tied during the fit, however we need to allow for different values of photon index between \textit{XMM-Newton}  and \textit{NuSTAR} in every observation. \textit{XMM-Newton} slopes are harder than the \textit{NuSTAR} derived ones and this discrepancy, likely due to residual inter-calibration issues, is, on average, of the order of $\Delta\Gamma \sim0.17$ \citep[see e.g.][and Appendix A for more details]{Capp16}. In this paper, we will report values of the photon index derived from \textit{NuSTAR} data.

	Following this procedure, we obtained a very good fit to the whole dataset, with $\chi^2$=3041 for 2765 d.o.f.~(see Fig. 2, bottom right side plot). The parameters of the hard X-ray components are fully compatible with those obtained from the fit of the data above 4 keV (Sect. 4.2 and Table 2).
	We report the best fit values for the parameters describing the soft-excess in Tab. 5.
	Most of the observed variability can be attributed to the \textit{nthcomp} normalization that varies among all the observations. On the other hand, the electron temperature is found consistent with being constant, while for the photon index marginal variations are observed, see Fig. 5, bottom right side plot.
	The measured warm corona temperature ($kT_{wc}$) is found to be, on average, $0.67\pm0.03$ KeV.
	
	The obtained electron temperature can be used to estimate the optical depth $\tau$ for the warm corona.
	Following \citet{Belo99}, and using his equation 13 and the average values for the $kT_{wc}$ and $\Gamma$ reported in Tab. 5, we estimate for the NGC~7469 warm corona $\tau_{\rm{wc}}$ to be $9.2\pm0.2$~.
	
	\begin{table}
		\caption{The best fit parameters for the soft excess, where $kT_{\rm{wc}}$ is the electron temperature of the warm corona. The obtained seed photons temperature is $\sim$3 eV with no signature of any variability among the observations. The best fit parameters for the hard X-ray model components are consistent, within the errors, with those obtained in the 4-78 keV band, and are not reported here for the sake of simplicity.}
	\centering
		\begin{tabular}{ c c c c c} 
			Obs &  $\Gamma_{\rm{soft}}$ & $kT_{\rm{wc}} (keV)$ &$\tau_{\rm{wc}}$ &Norm$_{nthcomp}$ ($10^{-3}$)  \\
			\hline
			\hline
			\\
			1 &$2.73^{+0.02}_{-0.02}$& $0.63^{+0.04}_{-0.03}$& 8.8$^{+0.5}_{-0.5}$ &$5.77^{+0.01}_{-0.01}$\\
			\\
			2 & $2.66^{+0.02}_{-0.02}$ & $0.71^{+0.04}_{-0.03}$ &8.7$^{+0.7}_{-0.6}$ & $5.42^{+0.01}_{-0.01}$\\
			\\
			3 & $2.62^{+0.02}_{-0.02}$&  $0.65^{+0.04}_{-0.03}$ &9.4$^{+0.8}_{-0.6}$ & $4.24^{+0.01}_{-0.01}$\\
			\\
			4 & $2.65^{+0.02}_{-0.02}$&  $0.68^{+0.03}_{-0.03}$ &9.0$^{+0.5}_{-0.4}$  & $4.10^{+0.01}_{-0.01}$\\
			\\
			5 & $2.58^{+0.02}_{-0.02}$&  $0.68^{+0.03}_{-0.05}$ &9.6$^{+0.3}_{-0.7}$  & $5.11^{+0.01}_{-0.01}$\\
			\\
			6 & $2.61^{+0.02}_{-0.02}$&  $0.65^{+0.04}_{-0.03}$ &9.5$^{+0.6}_{-0.6}$ & $4.56^{+0.01}_{-0.01}$\\
			\\
			7 & $2.62^{+0.02}_{-0.02}$ & $0.69^{+0.03}_{-0.03}$ &9.2$^{+0.6}_{-0.6}$   & $5.15^{+0.01}_{-0.01}$\\
			\hline

		\end{tabular}
	\end{table}

\begin{figure*}
	\includegraphics[width=0.5\linewidth]{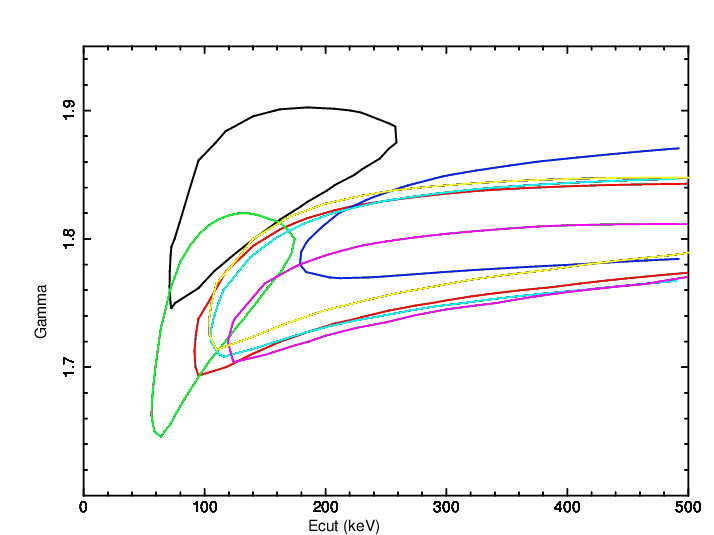}
	\includegraphics[width=0.5\linewidth]{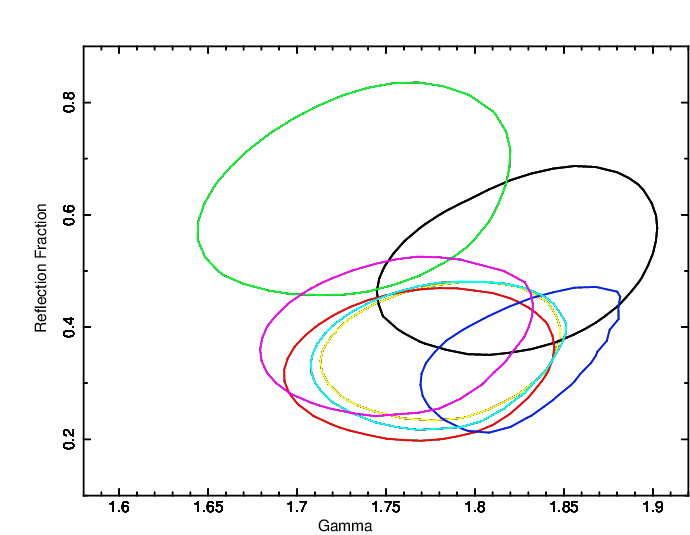}
	
	\includegraphics[width=0.5\linewidth]{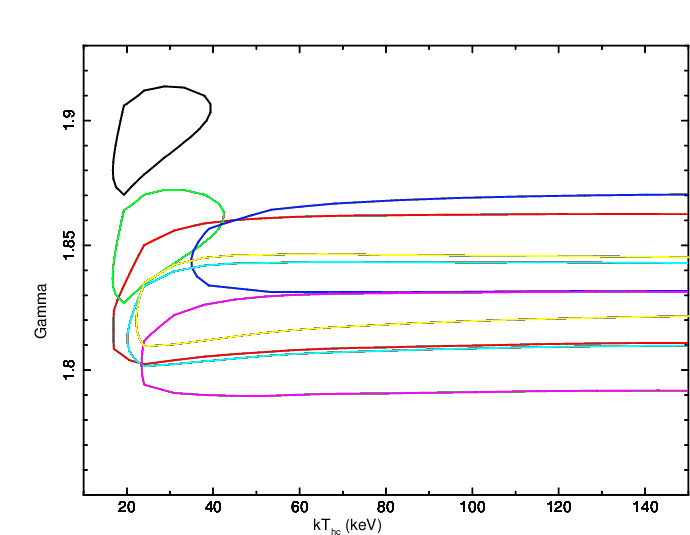}
	\includegraphics[width=0.5\linewidth]{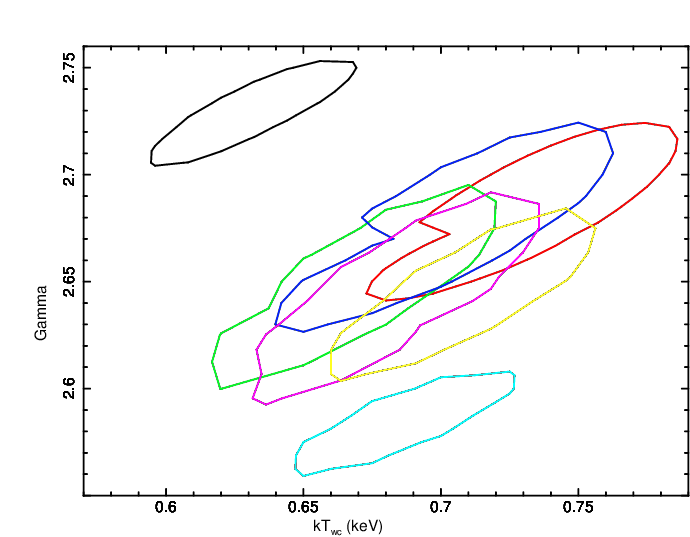}	
	\caption{$\;$The contours for the high energy cut-off and the photon index for all the \textit{NuSTAR} observations in the top left side panel. These contours account for the data analysed with \textit{xillver} and above 4 keV. The best fit parameters are displayed in Tab. 3.$\;$ The right top plot shows the best fit values for the reflection fraction as a function of the photon index for the different observations. The contours are computed on the \textit{NuSTAR} data above 4 keV adopting the model with \textit{xillver}.$\;$ On the bottom left hand figure the contour plots for the photon index and the hot electron temperature computed on the \textit{NuSTAR} data above 4 keV with \textit{xillvercp}.$\;$ The last plot shows the contour plots for the warm corona parameters for all the observations of our campaign.}\label{}
\end{figure*}
	
	\section{Discussion}

\subsection{A two-corona scenario}

The broad-band X-ray spectrum of NGC~7469 shows the presence of two main components, the primary power law at high energies, and a strong `soft excess' which starts dominating below $\sim4$ keV. This is commonly found in Seyfert galaxies \citep[e.g.][]{Pico05,Bian09,Scot12}. 

The high-energy X-ray spectrum can be phenomenologically characterized by a cut-off power law with average spectral index $\Gamma=1.78\pm0.02$ and high energy cut-off $E_{cut}$=$170^{+60}_{-40}$ keV. These parameters are consistent with being constant among all the observations, with only some marginal evidence of variability of the cut-off energy. The latter value is compatible within the errors with measures based on \textit{Suzaku} data \citep[$E_{cut}=119^{+65}_{-31}$ keV,][]{Patr11}, and \textit{BeppoSAX} \citep[$E_{cut}\sim150$ keV,][]{DeRo02}. On the other hand, the rapid $\Gamma$ variability reported by \cite{Nand00} could be due to the contamination from the soft excess, which could not be properly modelled in \textit{RXTE} data. Indeed, it is clear from our monitoring that the soft X-ray energy part of the spectrum varies more than the high energy part, generating variations of the hardness-ratio, which could mimic a photon index variability, if the two spectral components are not properly modelled separately (see e.g. Fig.\ref{lightcurves} and Fig. 2, top left hand panel.).
However, the rapid $\Gamma$ variability reported by \cite{Nand00} could also be due to a different state of NGC 7469 at the epoch of the \textit{IUE/RXTE} campaign. In fact, a comparison between our data and those studied by the authors reveals that the source was in a more variable state compared to 2015.

The cut-off power law which reproduces the high-energy spectrum of NGC~7469 can be naturally ascribed to Comptonization of the accretion disc photons onto a corona of hot electrons. Adopting a self-consistent Comptonization model, we recovered an average electron temperature of $kT_{\rm{hc}}=45^{+15}_{-12}$ keV and, under the assumption of a spherical geometry, an optical depth $\tau_{hc}=2.6\pm0.9$. These values are within ranges generally found in other Seyfert galaxies \citep[e.g.][]{Petr13,Petr17,Tort18a}, and are consistent with being constant among the observations of our monitoring campaign, in agreement with what found with phenomenological models. Interestingly, the coronal parameters measured in NGC~7469 lie along the $kT_e-\tau$ anti-correlation found by \citet{Tort18a} in a sample of Seyfert galaxies observed by \textit{NuSTAR}. As discussed in their paper, this anti-correlation is suggestive of variations in the heating/cooling ratio of the corona, as a result of different disk-corona geometries and/or intrinsic disc emission.
The soft excess cannot be satisfactorily modelled by simple phenomenological models, like a steep power law of a black body, as often found in large samples of objects and/or low signal-to-noise spectra \citep[][]{Bian09,Matt14,Ursi15}. Moreover, a self-consistent model in terms of blurred relativistic reflection is also statistically unacceptable for reproducing this soft excess.
 
On the other hand, an additional Comptonized spectral component provides a good representation of the soft excess in this source. Assuming the same seed photons as for the hot corona (i.e. those arising from the accretion disc), the temperature of the electron cloud responsible for the Comptonization of the spectrum is on average $kT_{wc}=0.67\pm0.03$ keV and the optical depth $\tau_{wc}=9.2\pm0.2$, again with marginal evidence for variability among the observations of our campaign. These values are well in agreement with those found in a sample of Seyfert galaxies, within the framework of the so-called two-corona model \citep[e.g.][]{Petr13,Mehd15,Roza15,Petr17}. According to this model, the soft X-rays emission is produced by Comptonization of the disc photons by a warm optically thick and extended medium (the warm corona), different from the compact, optically thin and hot medium (the hot corona), responsible for the high energy emission. The ranges of the coronal values found for the warm corona in Seyfert galaxies (including NGC~7469) are consistent with the warm corona covering a large fraction of a quasi-passive accretion disc, whose intrinsic emission is negligible, most of the accretion power being released in the warm corona itself \citep{Petr17}. 

\indent It is important to note that this two-corona scenario for NGC 7469 is also consistent with the observed optical/UV emission of this object by \textit{HST} and \textit{Swift} {UVOT}, as quantitatively shown by Mehdipour et al. accepted.
\subsection{The reprocessed components}

Along with the main continuum components discussed in the previous section, all the \textit{XMM-Newton} and \textit{NuSTAR} spectra of our campaign on NGC~7469 are characterized by the presence of a prominent emission line, readily identified as a neutral \ion{Fe} K$\alpha$ fluorescent line. We find that this line is unresolved, with no evidence of any broadening, and with a flux compatible to be constant among all the observations, as well as with the past \textit{XMM-Newton} observation in 2000 \citep{Blus03}. As expected if originating in Compton-thick matter, the iron line is associated with a reflection component, whose flux is also consistent with being constant among the observations of our campaign. Consequently, its reflection fraction with respect to the primary continuum is slightly variable ($R$ being in the range $0.3-0.6$), with higher values measured when the primary flux is lower. Self-consistent reflection models agree with a scenario where both the iron line and the reflection component arise from Compton-thick matter far from the accretion disc, as commonly found in the X-ray spectra of Seyfert galaxies \citep[e.g.][]{Bian09,Capp16,Alessia17}.

As already noted, we find no evidence for relativistic effects in the iron line profile of NGC~7469. However, it has to be compared with the rich literature for this source. \citet{Guai94} found the \ion{Fe} K$\alpha$ line to be narrow in a 40 ks \textit{ASCA} spectrum. In particular, the authors estimated the line-emitting region to be at several tens of Schwarzschild radii from the central engine of the galaxy. This result was confirmed by \cite{Nand97} with the same dataset. Subsequently, \citet{Blus03} working on \textit{XMM-Newton} data observed the \ion{Fe} K$\alpha$ line to be narrow, and they modelled it with a single narrow Gaussian line. On the other hand, the analysis of \textit{Bepposax} data made by \cite{DeRo02}, pointed to a relativistically broadened component of the line, together with an unresolved core. The presence of a relativistic component was confirmed, albeit marginally, by \textit{Suzaku} \citep{Patr11,Mant16}. 

Although we cannot exclude the presence of a broad component of the iron line in past observations, we may speculate that the narrow core we observe in \textit{XMM-Newton} data may be contaminated by other emission lines in spectra with lower spectral resolution and/or signal-to-noise ratio. Indeed, other emission features are clearly present in the \textit{XMM-Newton} spectra, as displayed in Fig.~\ref{iron}: a strong \ion{Fe}{xxvi} Ly$\alpha$ emission line, significantly detected in all the observations of our campaign, and weaker emission lines such as \ion{Fe}{xxv} He$\alpha$, neutral Fe K$\beta$ and Ni K$\alpha$, not always significant. While the latter two emission lines are expected to accompany the neutral \ion{Fe} K$\alpha$ emission, and therefore share the same origin, the other lines must arise in a much more ionised plasma. Such lines are often observed both in Seyfert 1s and in Seyfert 2s, and are likely produced in a Compton-thin, photoionized material illuminated by the nuclear continuum \citep[e.g.][]{Bianchi2002,Bianchi2005,Cons10}.

	\section{Conclusions}
	
In this paper we reported the spectral analysis of 7 simultaneous \textit{NuSTAR} and \textit{XMM-Newton} observations of the Seyfert galaxy NGC~7469 performed from June 2015 to December 2015 in the context of a multi-wavelength campaign.
	In the following we summarise the results of our analysis:
	\begin{itemize}
		\item NGC 7469 displayed a significant flux variability during this observational campaign, with intra-observation variability at few ks time-scales. We quantified this variability using the normalised excess variance estimator \citep[e.g][]{Pont12} $\sigma^2_{rms}$=0.0021$\pm$0.0005, which also allowed us to estimate the BH mass to be $M_{BH}=1.1\pm0.1\times 10^7$ $M_{\odot}$, in agreement with the measure based on reverberation mapping \citep{Pete14}.
		\item The high energy spectrum can be phenomenologically characterized by a cut-off power law with average spectral index $\Gamma=1.78\pm0.02$ and high energy cut-off $E_{cut}$=$170^{+60}_{-40}$ keV. These parameters are consistent with being constant among all the observations, with only some marginal evidence of variability of the cutoff energy. Using a realistic Comptonization model, the derived coronal parameters are $kT_{\rm{hc}}=45^{+15}_{-12}$ keV and $\tau_{\rm{hc}}=2.6\pm0.9$ for a spherical geometry.
		\item A strong soft-excess is observed in all the observations, extending up to 4 keV. The best description for this component is through another Comptonized spectrum, produced by a warm corona with $kT_{\rm{wc}}=0.67\pm0.03$ keV  and $\tau_{\rm{wc}}=9.2\pm0.2$, again with only marginal evidence for variability among the observations of our campaign. Indeed, most of the observed variability of the soft X-ray data may be simply ascribed to variations of the normalization of this component. The overall scenario is consistent with the so-called two-corona model \citep[e.g.][]{Petr13,Roza15,Petr17}, where  most of the accretion power is released in a warm optically thick and extended medium instead of the accretion disc. 
		\item A neutral \ion{Fe} K$\alpha$ emission line is present in all the observations. The line is found to be narrow, with no indications for relativistic broadening, and consistent with being constant. An accompanying Compton reflection component is also found to be constant among the observations, in agreement with a scenario where both components arise from Compton-thick matter located far away from the central BH. Weak neutral Fe K$\beta$ and Ni K$\alpha$ emission lines, only detected in some observations, have the same origin.		
		\item A \ion{Fe}{xxvi} Ly$\alpha$ emission line is significantly detected in all the observations of this campaign, and is likely to arise in a photoionized material illuminated by the central continuum, together with a weaker \ion{Fe}{xxv} He$\alpha$ emission line. 
			
\end{itemize}

	%%%%%%%%%%%%%%%%%%%%%%%%%%%%%%%%%%%%%%%%%%%%%%%%%%
	
	
\begin{acknowledgements}
We thank the referee for helping us in improving the quality of this paper.	
This work has made use of data from the \textit{NuSTAR} mission, a project led by the California Institute of  Technology,  managed  by  the  Jet  Propulsion  Laboratory, and funded by the National Aeronautics and Space Administration. We thank the \textit{NuSTAR} Operations, Software and Calibration teams for support with the execution and analysis of these observations. This research has made use of the \textit{nustardas} jointly developed by the ASI Science Data Center (ASDC, Italy) and the California Institute of Technology (USA). The work is also based on observations obtained with \textit{XMM–Newton}, an ESA science mission with instruments and contributions directly funded by ESA Member States and the USA (NASA).
RM and SB acknowledge financial support from the European Union Seventh Framework Programme (FP7/2007-2013) under grant agreement no. 312789. SB acknowledges financial support from the Italian Space Agency under grant ASI-INAF I/037/12/0.
POP acknowledges financial support from the CNES french agency and the CNRS PNHE.
GP acknowledges support by the Bundesministerium f$\ddot{u}$r Wirtschaft und Technologie/Deutsches Zentrum f$\ddot{u}$r Luftund Raumfahrt (BMWI/DLR, FKZ 50 OR 1408) and the Max Planck Society. SRON is supported financially by NWO, the Netherlands Organization for Scientific Research.BDM acknowledges support from the Polish National Science Center grant Polonez 2016/21/P/ST9/04025.The research at the Technion is supported by the I-CORE program of the Planning and Budgeting Committee (grant number 1937/12).
EB acknowledges funding from the European Union's Horizon 2020 research and innovation programme under the Marie Sklodowska-Curie grant agreement no. 655324. MC acknowledges financial contribution from the agreement ASI-INAF n.2017-14-H.O
\end{acknowledgements}
	
	\thispagestyle{empty}
	\bibliographystyle{aa}
	\bibliography{NGC7469_fin.bib}

	\appendix
	
	\section{$\Gamma$ discrepancy between \textit{XMM-Newton} and \textit{NuSTAR}}
	
	We reported in our analysis that our simultaneous \textit{XMM-Newton} and \textit{NuSTAR} data yielded photon indexes differing on average of $\sim0.17$. In this appendix, we report our investigations concerning a plausible origin for this issue.

\begin{itemize}

	\item \textit{Different energy band}: the energetic range of \textit{NuSTAR} is different with respect to the one of \textit{XMM-Newton}, thus we performed a simple analysis on the same observational band for both the satellites.
	For the whole sample of observations, we try to fit the spectra with a simple power law model in the common energy band 4-10 keV. The measurements for the photon indexes we obtained were still discrepant of the same amount.
	
	\item \textit{Pile-up}: our \textit{XMM-Newton} spectra could, at least marginally, suffer from pile-up. According to the XMM-Newton hand guide\footnote{$https://xmm-tools.cosmos.esa.int/external/\\xmm\_user\_support/documentation/uhb/epicmode.html$}, pile-up can affect pn observations performed in Small Window mode, when a total count-rate of $\sim25$ c/s is exceeded. Therefore, at least the first and the second \textit{XMM-Newton} observations could suffer pile-up problems. We thus tried to extract the spectra from source annular regions, using different values for the annulus inner radius (from 50 up to 200 pixels). However, our tests show that any choice of the inner annular radius improves the $\Delta\Gamma$ discrepancy between the \textit{XMM-Newton} and \textit{NuSTAR} spectra.
	
	\item \textit{Intrinsic variability}: \textit{XMM-Newton} observations are longer with respect to those performed by \textit{NuSTAR}, thus the difference in the two $\Gamma$ could be due to the not truly simultaneity of the observations. To verify this hypothesis we extracted \textit{XMM-Newton} spectra exactly in the same temporal range of those obtained using \textit{NuSTAR}\footnote{For the fourth observation the two satellites do not observe NGC~7469 simultaneously so that we cannot perform this test on it.}. As in the previous cases, analysing these truly simultaneous spectra does not affect the $\Delta\Gamma$ discrepancy.
	\end{itemize}
	
	We therefore conclude that the most likely origin for the spectral index discrepancy has to be found in residual inter-calibration issues between \textit{XMM-Newton} and \textit{NuSTAR}\footnote{A new SAS version was released when we were finalizing this paper (SAS version 16.1). We performed several tests to check if this could affect our results, but found that the reported $\Delta\Gamma$ between \textit{XMM-Newton} and \textit{NuSTAR} is not significantly improved.}, whose significativity may vary from observation to observation, and whose impact on the analysis is larger for high signal-to-noise data. We estimate that the results in this paper are not qualitatively affected by this issue, but a systematic uncertainty in the reported best-fit parameters should be taken into account.

	%%%%%%%%%%%%%%%%%%%%%%%%%%%%%%%%%%%%%%%%%%%%%%%%%%

	% Don't change these lines

%\begin{figure*}
%			\includegraphics[width=0.5\textwidth]{wall.png}					\includegraphics[width=0.5\textwidth]{wall.png}
%			
%			\includegraphics[width=0.5\textwidth]{wall.png}
%			\includegraphics[width=0.5\textwidth]{wall.png}
%\end{figure*}

\end{document}